\documentclass[12pt,a4paper]{article}
\pdfoutput=1
\usepackage{jheppub}
\usepackage{amsfonts}
\usepackage{bbm}
\usepackage{graphicx}
\usepackage[utf8]{inputenc}
\usepackage[T1]{fontenc}
\usepackage{amssymb,amsfonts,amsmath, mathtools, slashed}

\font\cmss=cmss10
\font\cmsss=cmss10 at 7pt
\font\manual=manfnt

\newcommand{\bi}{\begin{itemize}}
\newcommand{\ei}{\end{itemize}}

\newcommand{\bea}{\begin{eqnarray}}
\newcommand{\eea}{\end{eqnarray}}
\newcommand{\be}{\begin{equation}}
\newcommand{\ee}{\end{equation}}
\newcommand{\ben}{\begin{eqnarray*}}
\newcommand{\een}{\end{eqnarray*}}
\newcommand{\bem}{\begin{pmatrix}}
\newcommand{\eem}{\end{pmatrix}}
\newcommand{\bl}{\begin{align}}
\newcommand{\el}{\end{align}}
\newcommand{\beg}{\begin{gather}}
\newcommand{\eeg}{\end{gather}}

\newcommand{\cA}{\mathcal{A}}

\newcommand{\cI}{\mathcal{I}}

\newcommand{\cO}{\mathcal{O}}
\newcommand{\cS}{\mathcal{S}}

\newcommand{\IH}{\mathbb{H}}

\renewcommand{\a}{\alpha}

\renewcommand{\d}{\delta}
\newcommand{\e}{\epsilon}
               
\newcommand{\g}{\gamma}

\newcommand{\m}{\mu}
\newcommand{\n}{\nu}

\renewcommand{\o}{\omega}

\renewcommand{\t}{\tau}

\newcommand{\D}{\Delta}

\renewcommand{\O}{\Omega}

\newcommand{\1}{{\textbf{1}}}

\newcommand{\half}{\frac{1}{2}}

\newcommand{\Tr}{\mbox{Tr}}

\def\dbend{\lower3.5pt\hbox{\manual\char127}}

\def\IL{\relax{\rm I\kern-.18em L}}
\def\IH{\relax{\rm I\kern-.18em H}}
\def\rlx{\relax\leavevmode}

\def\ZZ{\rlx\leavevmode\ifmmode\mathchoice{\hbox{\cmss Z\kern-.4em Z}}
 {\hbox{\cmss Z\kern-.4em Z}}{\lower.9pt\hbox{\cmsss Z\kern-.36em Z}}
 {\lower1.2pt\hbox{\cmsss Z\kern-.36em Z}}\else{\cmss Z\kern-.4em
 Z}\fi}

\title{Quantum Slow Roll}

\author{Andr\'e Benevides}
\author{and}
\author{Atish Dabholkar}

\affiliation{International Centre for Theoretical Physics\\
Strada Costiera 11, Trieste 34151 Italy}

\abstract{We consider a scalar field coupled to massless fermions through Yukawa couplings, such as the Higgs field, in a Robertson-Walker spacetime. We compute the nonlocal quantum effective action as a functional of the background scalar field and the scale factor at one loop order by integrating the Weyl anomaly resulting from the fermions in the loop.  We show that the mode expansion of the  scalar field is modified by the quantum nonlocality and the power spectrum of scalar perturbations exiting the horizon exhibits a slow roll  with a red tilt proportional to the square of the Yukawa coupling even though the classical spacetime is exactly de Sitter. We comment on possible implications for cosmology.}

\makeatletter
\gdef\@fpheader{}
\makeatother

\begin{document}
\maketitle


\section{Introduction\label{sec:Intro}}

Loops of massless particles in curved spacetime result in a quantum effective action that depends non-locally on the background metric and other background fields. This  nonlocality of the effective action is a potential source of interesting quantum effects in cosmology or in  black hole physics and holography. Computing the 
  effective action is a well-posed problem in perturbation theory using the background field method, but is forbiddingly difficult to implement even at the one-loop level. One needs to compute  the heat kernel for the background-dependent quadratic fluctuation operator for all values of the proper time. This in general is not tractable. 

The  computation can be greatly simplified for Weyl-flat metrics for conformally coupled matter
\cite{Dabholkar:2015qhk, Bautista:2017enk} with Weyl-invariant classical action. In this case, under suitable conditions,  one can determine the quantum effective action by `integrating the Weyl anomaly'. Since the anomaly is a consequence of the short-distance regularization of the path integral, it is determined by the short proper-time expansion of the heat kernel which can be computed explicitly in a local expansion.  As a result, it is possible to determine \cite{Bautista:2017enk}  the quantum effective action in a closed form\footnote{We are interested in the consequences of the $\beta$-function anomalies away from a conformal fixed point and not in the conformal anomalies usually considered at the conformal fixed point, for example, the ones used to determine the Polyakov action or the Riegert action.} as we  review in \S\ref{sec:Weyl}. 

In four dimensions this method was implemented for the Yang-Mills field coupled to conformal matter \cite{Bautista:2017enk}.
The resulting action takes a simple form when expressed in terms of the conformal compensator $\Omega$ but has a rather complicated form when expressed covariantly in terms of the physical metric. One can check \cite{Bautista:2017enk} that it reduces\footnote{The complicated structure of the curvature expansion simplifies in a Weyl-flat spacetime because a large part of the curvature expansion evaluates to zero \cite{Barvinsky:1995it, Donoghue:2015nba}. This non trivial cancellation can be understood from the point of view of integrating the Weyl anomaly. The effective action can be decomposed into a Weyl-anomalous part and a Weyl-invariant part. The Weyl invariant part simplifies to the flat space action when the metric is Weyl-flat. One expects that this simplification must be present also in the covariant curvature expansion.} to the nonlocal  action to third order obtained using the Barvinsky-Vilkovisky covariant perturbation theory \cite{Barvinsky:1994hw,Barvinsky:1994cg,Barvinsky:1995it,Donoghue:2015nba} in the regime $R^{2} \ll \nabla^{2} R  $ for a typical curvature scale $R$. The advantage of our method for the class of  Weyl-flat metrics is that it has a greater range of validity effectively re-summing the covariant perturbation theory to all orders in curvatures. In particular, it is applicable also in  the opposite regime $R^{2} \gg \nabla^{2} R$, which is the relevant regime in a number of interesting situations in cosmology. For example, the quantum effective action for Maxwell gauge theory in Robertson-Walker spacetimes enables a systematic investigation of  scenarios for anomaly-induced primordial magnetogenesis \cite{Benevides:2018mwx} during inflation where the  curvatures vary much more slowly compared to the Hubble scale. 

The method of integrating the Weyl anomaly is not available when the classical theory is not Weyl invariant. In this case,  a different approach is required. For example, in quantum gravity near two dimensions with the cosmological term,  the classical action is not Weyl-invariant because the cosmological constant is dimensionful. It is nevertheless  possible to  determine the quantum effective action  \cite{Dabholkar:2015qhk, Bautista:2015wqy, Bautista:2019jau} by relying on the `fiducial Weyl invariance' discussed in \S\ref{sec:Fiducial} and the relation to Liouville theory. The  quantum effective action is much simpler in two dimensions than in four dimensions and  is instructive for gaining a better understanding of quantum nonlocality as discussed in more detail in \cite{Dabholkar:2015qhk, Bautista:2015wqy} and in  \S\ref{sec:KPZ}. 


The general lesson both from  Yang-Mills theory in four dimensions and  quantum gravity near two dimensions is that the effective coupling of different fields to  the metric can be different from the classical canonical coupling  because of nonlocal quantum effects. The relevant `effective metric'  depends upon the observable of interest and can be different from  the classical metric. 

In this paper, we examine another four-dimensional example where the nonlocal quantum effective action can be computed explicitly. We consider a (pseudo) scalar field in four dimensions coupled to $N_f$ massless fermions through Yukawa coupling, such as the Higgs field. Since the scalar may have mass and non-conformal coupling to the metric, the classical action is not Weyl invariant.  However, in the limit of large $N_f$, the quantum effective action is dominated by the fermionic contributions and the method of integrating the Weyl anomaly is still applicable. In \S\ref{sec:Scalar} we determine the resulting action by applying the method of \cite{Bautista:2017enk} described in \S\ref{sec:Weyl}. 

Light scalars and vectors with fermionic coupling have been of interest in various cosmological and astrophysical contexts. See, for example, \cite{Hu:2000ke, Arvanitaki:2021qlj}. Our considerations can be easily  extended to  complex scalar fields or to axionic fields and could be useful in these investigations. The quantum corrected dynamics of the scalar field during an inflationary phase is particularly interesting  because the quantum fluctuations of the scalar field can contribute to primordial density perturbations. With this motivation we examine in \S\ref{sec:Modes}  how  the mode expansion of the scalar field is altered because of the quantum nonlocality. Even though the quantum-modified equations of motion for the scalar are rather complicated, we are able to find analytic expressions for the modes at late times in terms of Airy functions.  In \S\ref{sec:Power} we compute the scalar power spectrum in de Sitter spacetime and  the spectral index\footnote{It was noted in \cite{Benevides:2018mwx} that the anomaly induced quantum fluctuations for a Maxwell field in an inflationary universe do not classicalize because the Maxwell action is classically Weyl invariant and the only source of coupling to the scale factor is through the Weyl anomaly. However the classical action of the scalar field considered in this paper is not Weyl invariant. As a result the quantum fluctuations of the scalar in an inflating universe do classicalize for the same reason as for an inflaton, but with the modifications induced by the quantum nonlocality.}. We find that,  even in exactly de Sitter spacetime, the power spectrum has a nonzero red tilt  that depends on the Yukawa coupling  and the number of fermions. Thus, the effective metric relevant for the scalar quantum fluctuations in this case is different from the classical de Sitter metric. 

 We expect that this general mechanism of `quantum slow roll' can have interesting applications in cosmology. In inflationary models, the inflaton plays a dual role as the field whose dynamics drives the nearly exponential expansion of the universe as well as the field that generates the primordial density perturbations. These two roles are logically separate. The classically slowly rolling potential required for these purposes is not natural from the perspective of effective field theory unless one can invoke some approximate global symmetry. By contrast, quantum slow roll is naturally slow in a weakly coupled theory without invoking artificially flat potentials. Thus, it may be interesting to consider scenarios where the two aspects of the inflationary cosmology are not necessarily linked. We have not considered here the re-entry of the scalar perturbations into the horizon after the end of inflation to produce density perturbations. It would be interesting to explore if this  mechanism of `quantum slow roll' can be embedded in specific cosmological scenarios.
   
\section{Quantum effective action from Weyl anomalies \label{sec:Weyl}}

Consider a classically Weyl invariant theory with classical lorentzian action $\cI_{0}[g, \chi_f]$ wick-rotated to  $\cS_{0}[g, \chi_f]$. The  local Weyl transformation of the  metric $g_{\m\n}$ is defined by
\be
\quad g_{\m\n} \rightarrow e^{2 \xi (x)} g_{\m\n} \, , \quad \,\, g^{\m\n} 
\rightarrow e^{-2 \xi (x)} g^{\m\n} \, , 
\ee
Other  fields denoted collectively as $\{\chi_{f}\}$ transform with their respective Weyl weights $\{\Delta_{f}\}$:
\be
\chi_{f}(x) \rightarrow e^{-\Delta_{f }\xi(x) }\chi_{f}(x)  \, .
\ee
In particular,  in four dimensions, a conformally coupled  scalar field has Weyl weight $1$, a fermion field has weight $3/2$, a gauge field has weight $0$ so that the kinetic terms are scale invariant. 

The local Weyl group $\mathcal{G}$ is an infinite dimensional  abelian group with generators $\{J_x$\} acting on the space of fields:
\be
J_x :=  -2 \, g^{\mu\nu}(x)\frac{\delta\quad}{\d g^{\m\n}(x)} - \Delta_{f}\, \chi_{f}(x)\frac{\d\quad }{\d \chi_{f}(x)}  \, . 
\ee
Treating the coordinate $x$ of the local scaling parameter $\xi(x)$ as a continuous index,  we can write an element  of this group as
\be
e^{\,\xi \cdot J} 
\ee
with the `summation' convention
\be
\xi \cdot J :=  \sum_x  \xi_x J_x := \int d^4x \,  \xi (x) J_x \, .
\ee
Weyl invariance of the classical action implies that
\be
J_{x }\left( \cS_{0}[g, \chi_{f}]\right) = 0 \, .
\ee

Consider now the one-loop  effective action $\cS = \cS_0 + \cS_1$, obtained by integrating the conformally coupled massless fields. Regularization of the path integral, for example, using short proper time cutoff of the heat kernel, is manifestly diffeomorphism invariant but not Weyl invariant.  Consequently, the Weyl-variation of the resulting action is nonzero. Since the lack of Weyl invariance is a result of a short distance cutoff, the anomalous Weyl-variation  is determined by the short proper time expansion  of the coincident heat kernel given by \cite{Bautista:2017enk}
\be\label{anomaly1}
J_{x } \left( \cS_1 [g, \chi_{f}]\right) := -\cA(x) \sqrt{g} = - 2\,\sum_{f} \,n_{f} \,\text{tr}\, K_{f}(x,x;\e) \sqrt{g}\,
\ee
where $n_{f}$ is a coefficient that differs for each type of field running in the loops. In particular  $n_{\psi}= -\half$ for a Weyl fermion $\psi$ and $n_{\varphi}= \half $ for a real scalar $\varphi$ \cite{Bautista:2017enk}.

The anomaly equation can be viewed as a first order differential equation on field space. If a metric 
$(g, \chi_{f})$ is on the Weyl-orbit of $(\bar g, \bar \chi_{f})$ with Weyl factor $\xi(x)$
\be\label{orbit}
(g, \chi_{f}) = e^{\xi \cdot J} (\bar g, \bar \chi_{f})\, .
\ee
then given the action $\cS_1 [\bar g, \bar \chi_{f}]$ one can compute the quantum effective action $\cS [g, \chi_{f}]$ by integrating the anomaly equation. 
It is convenient to implement this idea using the lemma discussed in \cite{Bautista:2017enk}. One starts with the trivial identity
\be
e^{\,\xi \cdot J}  = \textbf{1} + \int_{0}^{1} dt \,e^{\,t\, \xi \cdot J} \,\, \xi \cdot J \, .
\ee
to write\footnote{The argument $g$ of  the action functional $\cS[g, \chi_{f}]$ here refers to the covariant tensor $g_{\m\n}$ and not $g^{\m\n}$.} $\cS_1 [g, \chi_{f}]$ as 
\bea
 e^{\xi \cdot J} \left( \cS_1 [\bar g, \bar \chi_{f}] \right)  &=&  \left( \textbf{1}  + \int_{0}^{1} dt\,e^{t\, \xi \cdot J } \, \xi \cdot J \right)\left( \cS_1 [\bar g, \bar \chi_{f}]\right)\\
 &=&  \cS_1 [\bar g, \bar \chi_{f}]  + \int_{0}^{1} dt \,e^{t\, \xi \cdot J} \, \xi \cdot J \left(\cS_1 [\bar g, \bar \chi_{f}] \right)\\
 &=&  \cS_1 [\bar g, \bar \chi_{f}]  - \int_{0}^{1} dt \,e^{t\, \xi \cdot J} \left(\int d^4x \,\xi(x)\sqrt{g} \, \mathcal{A} [\bar g, \bar \chi_{f}] (x) \right) \, .
\eea
One concludes
\be\label{lemmaE}
\cS_1 [g, \chi_{f}] =  \cS_1 [\bar g, \bar \chi_{f}]  + \,  \cS_{\cA}[\bar g,  \bar \chi_{f}, \xi] \, ,
 \ee
 where 
 \be\label{lemmaE2}
 \cS_{\cA}[\bar g, \bar \chi_{f}, \xi] :=-\int_{0}^{1} dt \int d^4x \sqrt{\bar g \,e^{2\, t\, \xi(x)}} \,\, \xi (x) \,\, \mathcal{A} [\bar g \,e^{2\, t\, \xi}, \bar \chi_{f} \,e^{- \D_{f}\, t\, \xi}](x)
 \ee
is the anomalous contribution to the action. This lemma expresses the quantum effective action in terms of its Weyl variation as an integral over the Weyl orbit.

Since $\cS_{\cA}[\bar g, \bar \chi_{f}, \xi]$ is thus determined by the anomaly which is computable from the short-time expansion of the heat kernel, one can determine $\cS_1 [g, \chi_{f}]$ completely, if the `integration constant' $\cS_1 [\bar g, \bar\chi_{f}]$ can be computed. For a  Weyl-flat metric of the form
\be\label{weyl-flat}
g_{\m\n}(x) = e^{2\Omega(x)} \eta_{\m \n}
\ee
where $\eta_{\m \n}$ is the flat Minkowski metric one can choose $\bar g = \eta$ and $\xi(x) = \Omega(x)$.  
In this case 
our task is reduced to computing the quantum effective action $\cS[\eta, \bar \chi_f]$ with Minkowski metric which is often much simpler to evaluate. 

Lorentzian continuation of \eqref{lemmaE} gives a similar equation 
 \be\label{lemmaL}
\cI [g, \chi_{f}] =  \cI_0[g, \chi_f] + \cI_1 [\bar g, \bar \chi_{f}]  + \,  \cI_{\cA}[\bar g, \bar \chi_{f}, \xi] \, 
 \ee
 but with $\cI_{\cA}$ given by
 \be\label{lemmaL2}
 \cI_{\cA}[\bar g, \bar \chi_{f}, \xi] \coloneqq \int_{0}^{1} dt \int d^4x \sqrt{-\bar g \,e^{2\, t\, \xi(x)}} \,\, \xi (x) \,\, \mathcal{A} [\bar g \,e^{2\, t\, \xi}, \bar \chi_{f} \,e^{- \D_{f}\, t\, \xi}](x)
 \ee
 because the anomaly scalar does not change sign under Wick rotation. 
 
 By construction,  the effective action \eqref{lemmaE} can only depend on the physical  background fields  $g_{\mu\nu} = e^{2 \xi} \bar g_{\mu\nu}$ and $\chi_f = e^{-\Delta \xi} \bar \chi_f$. This implies that the effective action $\cS$ should be invariant a `fiducial' Weyl transformation of the fiducial metric $\bar g$ and other fields:
\be \label{fiducialE}
\xi \rightarrow \xi - \a(x)\ , \ \ \ \bar g_{\m\n} \rightarrow e^{2 \a (x)} \bar g_{\m\n}\ , \ \ \ \bar \chi_f \rightarrow e^{2 \a (x)} \bar \chi_f \,.
\ee 
Unlike the physical Weyl transformation corresponding to the Weyl transformation of the physical metric, fiducial Weyl transformation is a gauge transformation. It reflects the arbitrariness in splitting the physical metric $g$ into a fiducial metric $\bar g$ and the Weyl factor $\xi$. As a result, fiducial Weyl invariance must be respected as a gauge symmetry even in the quantum theory and cannot be anomalous. We further discuss the fiducial Weyl invariance of the action in \S\ref{sec:Fiducial}. 
 
\section{Scalar coupled to massless fermions\label{sec:Scalar}}

Consider the Euclidean action for a pseudo-scalar coupled to $N_f$ massless fermions:

\be
S_0 = \int_x \sqrt{g} \   \left[ \frac{1}{2} |\nabla \varphi|^2  + \frac{1}{2} m^2 \varphi^2 + \frac{\xi}{2} R \varphi^2 - \sum_{j= 1}^{N_f} \bar{\psi_j} \left( i\g^\mu \nabla_\mu + i y \varphi \g_5 \right) \psi_j\right] \, .
\ee
The action is invariant under parity under which the pseudoscalar is odd\footnote{To avoid notational confusion we denote the Yukawa coupling by $y$.}.  We choose to work with a pseudoscalar instead of a scalar so that linear and cubic terms are avoided in the effective action as a consequence of parity.
While the fermionic action is invariant under Weyl transformations, the scalar kinetic term is not. Hence, one can apply the lemma above to the fermionic contributions but not to  the scalar conbributions. 
This limitation can be circumvented by considering $N_f$ large so that the quantum effective action is dominated by the fermionic contributions and the error from ignoring the scalar contribution is small. 

The contribution from the the $N_f$ fermions at one-loop order is given by
\be \label{Log-Det}
N_f \log \det \left( \slashed{\nabla} + V \g_5\right) \ \,
\ee
where $V \equiv y \varphi$.
It is useful to rewrite \eqref{Log-Det} in terms of a quadratic differential operator. Notice that given $B \coloneqq  \slashed{\nabla} + V \g_5$
\be 
\log \det B = \half \log \left(\det B\right)^2 = \half \left(\log \det B + \log \det \g_5 B \g_5 \right) = \half \log \det \left(B \g_5 B \g_5 \right)
\ee
Therefore we can write
\be
\log \det \left( \slashed{\nabla} + V \g_5\right) = \frac{1}{2} \log \det \left( \slashed{\nabla} + V \g_5\right)\left( -\slashed{\nabla} + V \g_5\right) \, .
\ee
The one-loop contribution of $N_f$ fermions to the effective action can then be written as 
\be
\frac{N_f}{2} \log \det \left(  - \slashed{\nabla}^2 - \g_5 (\slashed{\nabla} V) - 2 V \g_5 \slashed{\nabla} + V^2 \1 \right) := \frac{N_f}{2} \log \det ( \cO)
\ee
with
\begin{equation}
	\cO :=  - D^2 + \left(-3 V^2 + \frac{R}{4}\right)\1 
\end{equation}
where  we have defined $D_\mu := \nabla_\mu + V \g_5 \g_\mu$ and used  $- \slashed{\nabla}^2 = -\nabla^2 + \frac{R}{4} \1$. 
The contribution to the anomaly from $N_f$ fermions can then be read off  from the heat kernel\footnote{The cohomology of the Weyl group on field space classifies the consistent terms that can appear as a Weyl anomaly and separates the non-trivial cohomological terms from the ones that are simply a variation of a local term in the action \cite{Bonora:1983ff}, such as $\nabla^2 R$ and $\nabla^2 \varphi^2$. In the model we consider, the only cohomologically non-trivial terms are the Euler density $E_4$, the square of the Weyl tensor $W^2$, the quartic potential $\varphi^4$ and the conformally coupled kinetic term $|\nabla \varphi|^2 + R \varphi^2/6$, consistent with what we have obtained from the heat kernel expansion. }
for the operator $\cO$ and is given by \cite{Gilkey:1995mj,Seeley:1967ea,Seeley:1969re,Hadamard:2014le,Minakshisundaram:1949xg,Minakshisundaram:1953xh,DeWitt:1965jb}
\be
\cA = \frac{N_f}{16\pi^2} \left[ \frac{11}{360} E_4 - \frac{1}{36} W^2 + 2 V^4 - \frac{1}{3} V^2 R -  2 \left( \nabla V \right)^2 \right]
\ee

The purely gravitational terms $E_4$ and $W^2$  are not presently of interest to us since they do not affect the mode expansion of the scalar field.
Each of the remaining terms can be integrated  separately to obtain the anomalous part of the effective action.
\begin{itemize}
	\item The $\varphi^4$ term is related to the running of the quartic scalar coupling. When integrated we obtain an expression that is a local  functional of $\bar g_{\m\n}$ and $\O$, but a non-local functional of $g_{\m\n}$. This non-local coupling to the metric is necessary because a running coupling implies the trace of momentum tensor is modified as a result of the Weyl anomaly. By integrating the $V^4$ term  one obtains
\be
\int_0^1 dt \sqrt{\bar{g}} e^{4 \O t} \bar{V}^4 e^{-4 \O t} \O = \sqrt{\bar{g}} \bar{V}^4  \O;
\ee
\item  The remaining two terms are the non-local effects corresponding to the kinetic term renormalization. The integration of such terms is slightly more convoluted since they involve derivatives of $\varphi$. Applying the lemma first to the $|\nabla V|^2$ term one obtains
\be
\int_0^1 dt \sqrt{\bar{g}} e^{4 \O t} e^{-2 \O t} \left[ \bar{\nabla} \left( e^{-\O t} \bar{V} \right)\right]^2 \O = \sqrt{|\bar{g}|} \O \left[ \left( \bar{\nabla} \bar{V} \right)^2 + \frac{1}{3} \bar{V}^2 \left( \bar{\nabla} \O \right)^2 - \frac{1}{2} \left( \bar{\nabla} \O . \bar{\nabla}\bar{V}^2 \right) \right];
\ee
Similarly the  $V^2 R$ term gives
\be
\int_0^1 dt \sqrt{\bar{g}} e^{4 \O t} \O e^{-2 \O t} \left[ \frac{R}{6} - \left( \bar{\nabla} \O \right)^2 t^2 - \bar{\nabla}^2 \O t\right] = \sqrt{|\bar{g}|} \O \bar{V}^2 \left[ \frac{R}{6} - \frac{1}{2} \bar{\nabla}^2 \O - \frac{1}{3} \left( \bar{\nabla} \O \right)^2  \right]
\ee
Combining the two and integrating by parts  one obtains
\be
\sqrt{\bar{g}} \left\{ \O \left[  \left( \bar{\nabla} \bar{V} \right)^2 + \frac{R}{6} \bar{V}^2 \right] + \frac{1}{2} \left( \bar{\nabla} \O \right)^2 \bar{V}^2 \right\}
\ee
\end{itemize}
Thus, the one-loop Lorenzian effective action is given by
\be \label{Effective_Action_Scalar}
\cI[g, \varphi] = \cI_0[g, \varphi] + \cI_1[\bar{g}, \bar{\varphi}] + \frac{N_f}{16\pi^2} \int_x \sqrt{-\bar g} \left\{ 2 \O \left[ \bar{V}^4 -  \left( (\bar{\nabla} \bar{V})^2 + \frac{\bar{R}}{6}\bar{V}^2 \right)\right] - \left(\bar{\nabla} \O\right)^2 \bar{V}^2\right\}
\ee
By substituting $\bar V = y e^{\O} \varphi$ and $\bar g_{\m \n} = e^{2 \O} g_{\m \n}$, the last term in \eqref{Effective_Action_Scalar} can be written as
\be
\frac{y^2 N_f}{16\pi^2} \int_x \sqrt{-g} \left\{ 2 \O \left[ V^4 -  \left( (\nabla V)^2 + \frac{R}{6}V^2 \right)\right] +  \left(\nabla \O\right)^2 V^2\right\}
\ee

The `integration constant' is the Weyl invariant part of the effective action. For  $\bar g_{\m \n} = \eta_{\m \n}$, it is given by
\be
\cI_1[\eta, \bar \varphi] = N_f \Tr \log \left[\slashed{\partial} + y  e^{\O} \varphi \g_5 \right]
\ee
which can be computed by Feynman diagrams in flat spacetime for rapidly fluctuating $\bar \varphi$.  
 More precisely, extra powers of $\bar \varphi ^2$ can be ignored compared to extra derivatives of $\bar \varphi$ if
 \be
 -\partial^2 \left(y^2e^{2\O}\varphi^2\right) \gg y^4e^{4\O}\varphi^4 .
 \ee
This is precisely the condition for the generalized curvature expansion \cite{Barvinsky:1994cg, Barvinsky:1995it, Barvinsky:2002uf, Barvinsky:2003rx}, where the trace of the logarithm involving the background field $\bar \varphi$ is expanded around the free theory with insertions of the generalized curvature $y^2 \bar \varphi^2$ up to the desired power. To make sense of the expansion of the operator in the logarithm around the free (derivative) operator, it is necessary that extra powers of derivatives are more important than extra powers of generalized curvatures.
\begin{figure}[!htbp]
\begin{center}
\includegraphics[width=0.7\linewidth]{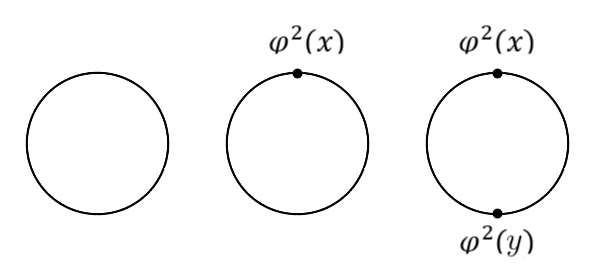}
\caption{Diagrams corresponding to each term in the expansion of the trace of the logarithm up to quartic order on $\varphi$.}
\label{diagrams}
\end{center}
\end{figure}
For (nearly) de Sitter metric this condition reads
\be 
-\partial^2 \left(\frac{\varphi^2}{H^2}\right) + 2\t \frac{\partial}{\partial \t}\left(\frac{\varphi^2}{H^2}\right) + 2 \left(\frac{\varphi^2}{H^2}\right) \gg y^2 \left(\frac{\varphi^4}{H^4}\right).
\ee
which can be satisfied either when $y^2 \left(\frac{\varphi^2}{H^2}\right) \ll 1$ or when the derivatives of $\left(\frac{\varphi^2}{H^2}\right)$ are large compared to its background value. This condition can be satisfied  as long as the occupation number is not too high. When $\varphi^2 \gg H^2$, the backreaction on the background metric be significant and one cannot regard the metric to be evolving independently of the $\varphi$ dynamics.

Under these conditions,  $\cI_1[\eta, \bar \varphi]$ can be evaluated using Feynman diagrams and it contains the flat space logarithms corresponding to each non-local  term in $\cI_\cA$: 
\bea
\cI_1[\eta, \bar \varphi] & \sim & 2\a \int \frac{d^3k}{(2\pi)^3}\frac{d\o}{2\pi}\  \bar \varphi_{\vec{k}}(\o) (k^2 - \o^2)\log \left(\frac{k^2 - \o^2}{\m^2}\right)\bar \varphi_{-\vec{k}}(-\o) +\cO(\bar \varphi^4) \\ \nonumber 
&=& 2\a \int_{\t \t' \vec{k} \o} a^2(\t)  \left[\varphi_{\vec{k}}^{''}(\t)+ 2 \frac{a'}{a} \varphi_{\vec{k}}'(\t) + \frac{a^{''}}{a}\varphi_{\vec{k}}(\t)\right] \varphi_{-\vec{k}}(\t')\ e^{i\o(\t-\t')} \frac{a(\t')}{a(\t)} \log \left(\frac{k^2 - \o^2}{\m^2}\right)
\eea
where we have ignored the $y^2\varphi^4$ terms, since we are only interested in the quadratic piece of the action to compute the mode expansion for $\varphi$. Here $\m$ is the  scale at which the scalar is canonically normalized, introduced in the process of renormalizing the UV divergences in $Z_{\varphi}$ and $Z_{\xi}$. The logarithms which depend on this renormalization scale  have the same coefficients as their non-local counterparts in \eqref{Effective_Action_Scalar} and a constant shift in $\O$ in \eqref{Effective_Action_Scalar} can  be compensated by a shift in $\log \m$.

To better understand the non-local logarithm, we  approximate the $\o$ Fourier integral by a delta function: 
\be 
\int \frac{d\o}{2\pi} e^{i\o(\t-\t')}\log \left(\frac{k^2 - \o^2}{\m^2}\right) \sim \d(t-t')\log \left(\frac{k^2}{\m^2}\right) \, .
\ee
In this approximation
\bea
\cI_1[\eta, \bar \varphi] & \sim & 2\a \int_{\t \vec{k}} a^2(\t)  \left[\varphi_{\vec{k}}^{''}(\t)+ 2 \frac{a'}{a} \varphi_{\vec{k}}'(\t) + \left( k^2 + \frac{a^{''}}{a}\right)\varphi_{\vec{k}}(\t)\right] \varphi_{-\vec{k}}(\t)\ \log \left(\frac{k^2}{\m^2}\right)  \\ \nonumber 
&=& 2\a \int_{\t \vec{k}} a^4(\t)  \left[- a^{-2}\varphi_{\vec{k}}'\varphi_{-\vec{k}}' + k^2a^{-2}\varphi_{\vec{k}}\varphi_{-\vec{k}} + \frac{R}{6}\varphi_{\vec{k}}\varphi_{-\vec{k}}\right] \ \log \left(\frac{k^2}{\m^2}\right)
\eea
which can be seen as the logarithmic running of the kinetic  and the $\xi R \varphi^2$ term as  $k^2$ is varied. 

It would be interesting to reproduce our conclusions by a computation of one-loop Feynman diagrams in de Sitter spacetime. In particular, it would be important to understand better the breakdown of time-translation symmetry.   There are a number of subtleties in global de Sitter spacetime  \cite{Allen:1985ux, Allen:1987tz} which are a consequence of the special symmetries that exist only when the metric is  exactly de Sitter. For example, all curvatures are constant for the exact de Sitter metric, and the usual expressions for the scale factor in terms of $g_{\m\n}$ become IR divergent. It is also well-known that zero modes in global de Sitter can complicate the analysis and can lead to the breakdown of de Sitter symmetry.  We have side-stepped these issues here  by considering a more general Robertson-Walker metric which  does not have all the symmetries of global de Sitter and approaches de Sitter as a limit.

\section{Mode Expansion \label{sec:Modes}}

The mode expansion of the field $\varphi$  is modified by the nonlocal coupling to the scale factor $\Omega$. The relevant quadratic Lorentzian effective action for determining the modes is given by
\bea
\cI[g, \varphi] &=& -\frac{1}{2}\int d^4x \sqrt{- g} \left[ I^2(\O)|\nabla \varphi |^{2} + \frac{2\a}{3} \O R\, \varphi^{2} -2 \a |\nabla \O |^{2} \varphi^{2} + m^2 \varphi^2 + \xi R \varphi^2 \right] \\
&=& \frac{1}{2}\int d^4x\ a^2 I^2(\O) \left[ (\varphi')^2 - |\vec{\nabla}\varphi |^2 - \left(\frac{m^2a^2}{I^2} + \frac{a''}{a}\left(1 + \frac{6\xi - 1}{I^2}\right) + \frac{2\a}{I^2} \left(\frac{a'}{a}\right)^2\right)\varphi^2\right]
\nonumber
\eea
where $I^2(\O) \coloneqq 1 + 4 \a \left(\O - \log(k/\m)\right)$, with $\a \coloneqq \frac{y^2  N_f}{16\pi^2}$. For $\chi \coloneqq aI\varphi$, the action takes the form
\be
\cI = \frac{1}{2}\int d^4x \left[(\chi')^2 - |\vec{\nabla}\chi |^2 - m^{2}_{eff}(\t)\chi^2 \right]
\ee
where the effective time dependent mass is given by
\be
m^{2}_{eff}(\t) \sim m^2 a^2 \left(1 - 4\a \log \frac{a H}{k}\right) - \frac{a''}{a}\left[2\a + (1-6\xi)\left(1 - 4\a \log \frac{a H}{k}\right)\right]
\ee

One can expand the field $\chi$ in terms of Fourier modes using translation invariance 
\be 
\chi(\vec{x}, \t) = \frac{1}{\sqrt{2}} \int \frac{d^3k}{(2\pi)^3}\left[a_{\vec{k}} v_k^{*}(\t) + a_{\vec{k}}^{\dagger} v_k(\t) \right]e^{i \vec{k}. \vec{x}} \, .
\ee
The mode functions $v_k(\t)$ satisfy the equation of motion 
\be
v_k^{''}(\t) + [k^2 + m_{eff}^2(\t)]\ v_k(\t) = 0.
\ee
In particular, in de Sitter spacetime with $a(\t) = \exp [\Omega(\t)] = -\frac{1}{H\t}$, one obtains
\be 
m_{eff}^2(\t) = \frac{1}{\t^2}\left\{\frac{m^2}{H^2}+ 12 \xi - 2 + 4\a \left[\left( \frac{m^2}{H^2} + 12\xi - 2\right)\log\left(\frac{H|\t|k}{\m}\right)- 1\right]\right\}.
\ee
To avoid large logarithms from RG running, a natural choice for the renormalization scale is $\m = H$ so that the scalar is canonically normalized at the scale of the de Sitter horizon.

At early times,  $k^2$ dominates over $m_{eff}^2(\t)$ and the two independent solutions are asymptotically proportional to $e^{\pm i k \t}$. One can choose the Bunch-Davies vacuum by selecting
\be \label{BD-vacuum}
v_k(\t) = \frac{1}{\sqrt{k}}e^{i k \t}.
\ee

To analyze the late time solution, it is useful to change variables from $\t$ to $N \coloneqq \log(a H/k)$, to write the equation of motion as
\be 
\frac{d^2 v_k}{dN^2} + \frac{d v_k}{dN} + \left\{e^{-2N} + \frac{m^2}{H^2} + 12 \xi - 2 - 4\a\left[\left( \frac{m^2}{H^2} + 12\xi - 2\right)N + 1\right]\right\}v_k = 0 .
\ee
Deep outside of the horizon,  one can neglect the $e^{-2N}$ term to obtain
\be \label{mode-equation}
\frac{d^2 v_k}{dN^2} + \frac{d v_k}{dN} + \left\{ \frac{m^2}{H^2} + 12 \xi - 2 - 4\a\left[\left( \frac{m^2}{H^2} + 12 \xi - 2\right)N + 1\right]\right\}v_k = 0 .
\ee
For comparison, recall that when $\a = 0$, one has the standard late time solution
\be \label{Polynomial_Solution}
v_k(N) = e^{-N/2}\left[ \tilde{C}_k\  e^{-\n N} + \tilde{D}_k\  e^{\n N}\right] =  C_k\  a^{-\left(\n+\frac{1}{2}\right)} + D_k\  a^{\left(\n-\frac{1}{2}\right)},
\ee
where $\n \coloneqq \sqrt{\frac{9}{4} - \frac{m^2}{H^2} - 12\xi}$ and $C_k$ and $D_k$ are constants.
At late times $D_k a^{\left(\n-\frac{1}{2}\right)}$ dominates. Matching solutions at the time of horizon crossing $|k\t| = 1$, results in the late time solution
\be \label{Late_Time_Vk}
v_k(\t) \sim \frac{1}{\sqrt{k}}\left(\frac{H a(\t)}{k}\right)^{\n - \frac{1}{2}} = \frac{1}{\sqrt{k}}|k\t|^{\frac{1}{2}-\n}.
\ee

With $\a \ne 0$, the nonlocal effects are turned on. The mode equation \eqref{mode-equation} now has a linear term proportional to $N$ which cannot be absorbed in the redefinition of the mass or of $\xi$. It is nevertheless possible to find an analytic solution at late times in terms of the two independent Airy functions $A_i(z)$ and $B_i(z)$:
\be \label{Airy_Solution}
v_k(N) = e^{-N/2}\left[ C_k\ A_i(z) + D_k\ B_i(z)\right].
\ee
where $C_k$ and $D_k$ are constants and
\be 
z \coloneqq 2^{-4/3} \frac{\n^2 + 16\a + 4\a\left( \frac{m^2}{H^2} + 12 \xi - 2\right)N}{(\a)^{2/3} \left( \frac{m^2}{H^2} + 12 \xi - 2\right)^{2/3}} \ \ , \ \ \ \ \n \coloneqq \sqrt{\frac{9}{4} - \frac{m^2}{H^2} - 12\xi}
\ee
In the $\a \rightarrow 0$ limit the combination of Airy functions \eqref{Airy_Solution} reduces to the polynomial solution \eqref{Polynomial_Solution}, up to constants that can be reabsorbed in the definition of $C_k$ and $D_k$.
At Late times ($N \gg 1$) $B_i(z)$ dominates and has the small $\a$ expansion:
\be 
B_i(N) \propto \ \exp\left[ \left(\n + \frac{8\a}{\n}\right)N + \frac{\a b}{\n}N^2\right], 
\ee
where $b \coloneqq 2 + \frac{m^2}{H^2} - 12 \xi$.
Matching the solution at  horizon crossing $|k\t|=1$ leads to
\be 
v_k \sim \frac{1}{\sqrt{k}} \left|k\t \right|^{\half - \n -\frac{8\a}{\n}} \exp \left[ \frac{\a b}{\n}\left(\log|k\t|\right)^2\right].
\ee

\section{Quantum red tilt in the power spectrum\label{sec:Power}}
The power spectrum $\D^2(k)$ for the field $\varphi_k = v_k/(a I)$ is given by
\be \label{Power_Spectrum}
\D^2(k)  = \frac{k^3 |v_k|^2}{\left(2\pi a I\right)^2} .
\ee
Before horizon crossing the modes are  oscillatory functions as in \eqref{BD-vacuum} with power spectrum
\be 
\D^2(k) \sim  \frac{1}{4\pi^2} \left(\frac{k}{a}\right)^2 \left[1-4\a \log \frac{aH}{k}\right]  = \frac{H^2}{4\pi^2} \left|k\t\right|^2 \left(1+4\a \log|k\t|\right).
\ee
Note that as a function of the physical wavelength $L_{ph} = 2\pi \left(k/a\right)^{-1} = 2\pi/(H k|\t|)$, the spectrum before horizon crossing grows as
\be
\D^2(k) \sim L_{ph}^{-2} \left[1 -4\a \log\left(\frac{H L_{ph}}{2\pi}\right)\right]
\ee
which indicates a slightly faster decay compared to $\a = 0$. 

For the power spectrum after horizon crossing, consider first the unperturbed case $\a = 0$.
In this case the mode function is given approximately by \eqref{Late_Time_Vk}. Thus, 
\be 
\D^2(k)  = 
\frac{H^2}{4\pi^2} \left|k\t\right|^{3-2\n} \qquad\qquad ( {\a = 0})
\ee
In the massless, minimally coupled limit, $m^2 = 0$ , $ \xi = 0$ and therefore $\n = 3/2$. The unperturbed spectrum is exactly flat as expected:
\be 
\D^2(k)  = \frac{H^2}{4\pi^2} \qquad\qquad ( {\a = 0}, \ {m^2 = 0}, \ {\xi = 0}).
\ee

Let us now include the non-local effects by turning $\a \ne 0$.
\be \label{Scalar_Power_Spectrum}
\D^2(k) = \frac{H^2}{4\pi^2} \left|k\t\right|^{3-2\n -\frac{16\a}{\n}}
\left(1+ 4\a\log |k\t|\right) \exp\left[\frac{2\a b}{\n} \left(\log |k\t| \right)^2 \right].
\ee
Notice that \eqref{Scalar_Power_Spectrum} is a slowly decaying function as $\t \rightarrow 0$ (in future) for $\frac{22\a}{9}\lesssim \n \lesssim \frac{3}{2} - \frac{4\a}{9}$. 
The massless minimally coupled scalar is slighly outside of that range. Indeed for $m^2 = 0$ and $\xi =0$ one obtains the following slowly growing function
\bea
\D^2(k)  &=& \frac{H^2}{4\pi^2} \left|k\t\right|^{-\frac{32\a}{3}} \left(1+ 4\a \log |k\t| \right) \exp\left[\frac{8\a}{3}  \left(\log |k\t| \right)^2 )\right] \nonumber \\ 
&\sim & \frac{H^2}{4\pi^2} \left|k\t\right|^{-\frac{20\a}{3}}\exp\left[\frac{8\a}{3} \left(\log |k\t| \right)^2\right].
\eea
In terms of the physical wavelength $\D^2(k)$ scales as
\be
\D^2(k) \sim \left(\frac{H}{2\pi} \right)^2 \left(\frac{H L_{ph}}{2\pi}\right)^{\frac{20\a}{3}} \exp\left[\frac{8\a}{3} \left(\log \frac{HL_{ph}}{2\pi} \right)^2\right].
\ee
Perturbation theory breaks down at very late times because of large logarithms and one cannot trust the solution anymore. However, the  modes will have exited the horizon way before this happens.


The spectral index $n_s$ is defined as
\be
n_s = 1 + \frac{d \log \left(\D^2(k) \right)}{d \log k}.
\ee
It is convenient to cast the power spectrum \eqref{Scalar_Power_Spectrum} as
\be
\D^2(k) = \frac{H^2}{4\pi^2} \exp\left[\left(3-2\n -\frac{16\a}{\n} + 4\a \right) \log |k\t|  + \frac{2\a b}{\n} \left( \log |k\t| \right)^2\right].
\ee
It then follows that
\be 
n_s = 1 + 3-2\n + \a\left(4 - \frac{16}{\n}  + \frac{4b}{\n}\log |k\t| \right).
\ee
In particular, in the $m^2 = 0$ and $\xi = 0$ case, one obtains
\be 
 n_s = 1 - \a\left(\frac{20}{3} - \frac{16}{3}\log |k\t| \right).
\ee

In summary, the power spectrum at horizon exit for a massless scalar is not flat even in exact de Sitter spacetime. Instead, it exhibits a `quantum slow roll' induced by the quantum nonlocality which mimics the power spectrum of a classically slowly-rolling scalar field such as an inflaton.
The first part of the effect mimics the $\a = 0$ case with a mass $\frac{m^2}{H^2} =  10\a - \frac{100\a^2}{9}$. The second part is quite distinct and has a characteristic logarithmic growth.

\appendix

\section{Quantum nonlocality and fractal geometry in two dimensions \label{sec:KPZ}}

Consider Einstein gravity in $2+\e$ dimensions \cite{Bautista:2015wqy}, with metric $g_{\m\n} = e^{2\O} h_{\m\n}$,     gravitational coupling $Q^{-1}$ and cosmological constant $\m$. In terms of $h_{\m\n}$ and $\O$ the Euclidean action reads:
\be
\cS_0[\O, h] = \frac{Q^2}{4\pi}\int d^{2 + \e} \sqrt{h} \left[\frac{R_h}{\e} + |\nabla_h \O|^2 +  R_h \O - \frac{4\pi\m}{Q^2} e^{2\O}\right] \, . \label{Liouville-action}
\ee
Apart from the Einstein term, this is the classical action of spacelike\footnote{The Einstein-Hilbert action near two dimensions with a conformal compensator leads to timelike Liouville theory  \cite{Bautista:2015wqy} which can be analytically continued to spacelike Liouville. We consider spacelike Liouville to more easily connect with KPZ scaling but our considerations can be easily extended  to the timelike case.}  Liouville field $\phi$ with $\phi = Q\O$. The Lorentzian action can be obtained by analytic continuation \cite{Bautista:2015wqy}. 

In the quantum theory, it is well-known that the 'cosmological constant' operator   acquires an anomalous dimension  \cite{Knizhnik:1988ak, Distler:1988jt,  Kawai:1992np, Kawai:1993fq, David:1988hj}. 
As a result the quantum effective action is given by
\be \label{Quantum_Liouville}
\cS[\O, h] = \frac{Q^2}{4\pi}\int d^{2 + \e} \sqrt{h} \left[\frac{R_h}{\e} + |\nabla_h \O|^2 + R_h \O - \frac{4\pi\m}{Q^2} e^{2 Q b \O}\right]
\ee
where $b$ is defined implicitly by 
\be
Q = \frac{1}{b} + b \qquad or\qquad  Qb = 1 + b^2 .
\ee
Classically $Qb = 1$, the quantum correction proportional to $b^2$ can be ignored, and one recovers the classical action \eqref{Liouville-action}. 

This anomalous `gravitational dressing' of the cosmological constant operator is related to the fractal geometry in two dimensions. See \cite{Ginsparg:1993is} for a review. The  KPZ critical exponent $\g_{str }$ for  the fixed area partition function is defined by 
\be 
Z[A] = \int D\phi\ e^{-S[\O, h]} \d \left(A - \int d^2x\sqrt{h}e^{2b\phi}\right)  \sim A^{\half \left(\g_{str}-2\right)\chi-1}
\ee
in the large area limit $A \rightarrow \infty$ 
 where $\chi = \frac{1}{4\pi}\int d^2x R$ is Euler character. 
As we see below, the anomalous KPZ scaling is intrinsically related to non-local terms in the action. However, one can shift the non-locality from one term to the other depending on the choice of the metric. 

One choice for the metric is $G_{\m\n}$ defined by 
\be 
G_{\m\n} = e^{2Qb\O} h_{\m\n}
\ee
so that the cosmological term is local. 
In terms of $G_{\m\n}$ the effective action is
\be 
\cS = \frac{Q^2}{4\pi} \int d^{2 + \e} \sqrt{G} \left[\frac{R_G}{\e} - \frac{b}{Q} R_G \Sigma_G  + \frac{b^2}{Q^2} |\nabla \Sigma_G |^2 - \frac{4\pi\m}{Q^2}\right]
\ee
where $\Sigma_G \coloneqq 2Qb\O$. Thus, the Einstein term is nonlocal because $\Sigma_G$ cannot be represented as a local functional of $G_{\m\n}$. 
To obtain the KPZ scaling in this metric frame one  performs the scaling  $G_{\m\n} \rightarrow e^{2\xi}\ G_{\m\n}$which implies $\Sigma_G \rightarrow \Sigma_G + \xi$, with $\xi = \half \log A$. Under this scaling the terms in the action shift as
\bea 
& & \frac{Q^2}{4\pi\e} \int d^{2+\e}x\sqrt{G}\ R_G \ \rightarrow \ \frac{Q^2}{4\pi\e} \int d^{2+\e}x\sqrt{G}\ R_G  + Q^2 \xi \chi \\ 
& & \frac{Q b}{4\pi }\int d^{2+\e}x\sqrt{G}\ R_G \Sigma_G \ \rightarrow \ \frac{Q b}{4\pi }\int d^{2+\e}x\sqrt{G}\ R_G \Sigma_G + Q b \xi \chi
\eea
The total action scales as
\be 
\cS \rightarrow \cS + \left(Q^2 -Qb \right)\xi \chi = \cS + \frac{Q}{b}\xi \chi
\ee
which in turn implies the KPZ scaling of the fixed area partition function \be \label{Fixed_Area_Scaling}
Z(A) = e^{-\half \chi \frac{Q}{b} - 1}Z(1) \qquad with \qquad  \g_{str} = 2 - \frac{Q}{b}
\ee

Another choice for the metric is the original classical metric $g_{\m\n}$:
 \be
 g_{\m\n} = e^{2\O}h_{\m\n} \, .
 \ee
In this frame,  the effective action is
\be 
\cS = \frac{Q^2}{4\pi} \int d^{2 + \e}x \, \sqrt{g} \left[\frac{R_g}{\e} + \frac{4\pi\m}{Q^2}e^{2(bQ -1)\Sigma_g}\right]
\ee
where $\Sigma_g =\O$.
Now, the Einstein term is local but the cosmological term is nonlocal. 
One can check that the Einstein-Hilbert term scales as
\be 
\frac{Q^2}{4\pi} \int d^{2+\e}x \, \frac{R_g}{\e}\sqrt{g} \rightarrow \frac{Q^2}{4\pi} \int d^{2+\e}x \,\frac{R_g}{\e}\sqrt{g} e^{\frac{\e\xi}{Qb}} = \frac{Q^2}{4\pi} \int d^{2+\e}x \,\frac{R_g}{\e}\sqrt{g} + \frac{Q}{b}\xi \chi 
\ee
and one reproduces the KPZ scaling \eqref{Fixed_Area_Scaling}.

Both metric frames reproduce the same KPZ exponents, since they are related by a (nonlocal) field redefinition. However,  the semiclassical solutions  in terms of the two metrics are different: upon Lorentzian continuation, the metric $G_{\m\n}$ corresponds to exact de Sitter spacetime whereas the metric $g_{\m\n}$ corresponds to a slowly decaying quasi de Sitter spacetime \cite{Bautista:2015wqy}. 

Which of these metrics is relevant is determined by the observable of interest. For example, if one introduces additional matter fields such as a fermion $\psi$, then  composite operators  have nonlocal couplings to the metric determined by their conformal dimensions. For example, the gravitationally dressed mass term is given by \cite{Ishimoto:2005ag}
\begin{equation}
	\int d^{2 +\e} x \sqrt{h} \, \bar  \psi \psi e^{a Q \O} \, .
\end{equation}
where $a$ in the quantum theory is different from its classical value $1/2 Q$ and has an anomalous quantum correction. Thus, the matter fields  effectively couple to a metric different from both $G_{\m\n}$ and $g_{\m\n}$ and is not exactly de Sitter. This is closely analogous to the effective metric seen by the scalar field $\varphi$ in four dimensions as we have analyzed in \S\ref{sec:Scalar}. 

\section{Fiducial Weyl invariance \label{sec:Fiducial}}

 The integrated anomaly \eqref{lemmaE2} is not manifestly invariant under fiducial Weyl transformations. We now show that it is indeed invariant by using the structure of the Weyl group.
Applying the transformation \eqref{fiducialE} to \eqref{lemmaE2} one obtains
\be
\cS'[g, \chi_f] = \cS[e^{2\a} \bar g, e^{-\D \a} \bar \chi_f] + \cS_{\cA}[e^{2\a} \bar g, e^{-\D \a} \bar \chi_{f},  \xi - \a] \, .
\ee
The term  $\cS[e^{2\a} \bar g, e^{-\D \a}\chi_f]$ is just the Weyl transformation of $\cS[\bar g, \bar \chi_f]$ with parameter $\a$, for which one can apply the same lemma again to obatin
\be
\cS'[g, \chi_f] = \cS[\bar g, \bar \chi_f] + \cS_{\cA}[\bar g, \bar \chi_{f},  \a] + \cS_{\cA}[e^{2\a} \bar g, e^{-\D \a} \bar \chi_{f},  \xi - \a]
\ee
which in turn should be equal to $\cS[g, \chi_f]$. This implies the condition\footnote{The Wess-Zumino consistency condition \cite{Wess:1971yu} can be seen as an infinitesimal form of this condition.}
\be \label{fiducialC}
\cS[g, \chi_f] - \cS'[g, \chi_f] = \cS_{\cA}[\bar g, \bar \chi_{f},  \xi] - \cS_{\cA}[\bar g, \bar \chi_{f},  \a] - \cS_{\cA}[e^{2\a} \bar g, e^{-\D \a} \bar \chi_{f},  \xi - \a] = 0 \, .
\ee
It is useful to change variables in the $t$ integral from $t$ to $u = e^{at + b}$ to obtain
\be \label{changeofvariables}
\int_0^1 dt\ a\ e^{at + b}\ f(t) \ =\ \frac{1}{d} \int_{e^b}^{e^{a + b}} du\ f\left(\frac{1}{d} \log u\right) \, .
\ee
To simplify the notation let us define the following scaled quantities
\be\label{scaledquantities}
g_{\xi t} \equiv  e^{2 \xi t} \bar g \ , \ \ \ \chi_{\xi t} \equiv  e^{-\D \xi t}\ , \ \ \ \cA_{\xi t} \equiv \cA[g_{\xi t}, \chi_{\xi t}] \, .
\ee

Applying \eqref{changeofvariables} to \eqref{fiducialC} one can write
\bea
\cS_{\cA}[g_\a, \chi_{f\, \a},  \xi - \a] &=& - \int d^dx \sqrt{\bar g}\int_0^1 dt\ (\xi - \a)\ e^{d(\xi - \a)t + d\a}\ \cA_{(\xi - \a)t + \a}  \\
&=& - \frac{1}{d} \int d^dx \sqrt{\bar g }\int_{e^{d\a}}^{d\xi} du\ \cA_{\frac{1}{d} \log u} \, .
\eea
On the other hand,
\bea
\cS_{\cA}[\bar g, \bar \chi_{f},  \xi] - \cS_{\cA}[\bar g, \bar \chi_{f},  \a] &=& - \int d^dx \sqrt{\bar g }\  \int_0^1 dt\ \left(e^{d\xi t}\xi \cA_{\xi t} - e^{d\a t}\xi \cA_{\a t} \right)  \\
&=& - \frac{1}{d} \int d^dx \sqrt{\bar g }\int_{e^{d\a}}^{d\xi} du\ \cA_{\frac{1}{d} \log u} = \cS_{\cA}[g_\a, \chi_{f\, \a},  \xi - \a] \, .
\eea
Thus, the condition for fiducial Weyl invariance \eqref{fiducialC} is indeed satisfied.

The particular choice of the Minkowski metric as the fiducial metric,  $\bar g_{\m\n} = \eta_{\m\n}$, is arbitrary and does not affect the final result as we have seen above. With this particular choice, the Weyl parameter is the conformal factor of the metric $\O$. An advantage of this choice is that the `integration constant' in the lemma discussed in \S\ref{sec:Weyl} can be evaluated relatively easily using flat space Feynman diagrams.  
Another interesting choice for the fiducial metric could be the de Sitter metric especially if one is interested in analyzing the dynamics in quasi de Sitter spacetime as in classically slowly rolling inflation. In this which the Weyl parameter would be the conformal factor of the metric relative to the de Sitter metric, that is $\O -\O_{dS}$. For quasi de Sitter space, this difference would be small.

\subsection*{Acknowledgments}

We thank Teresa Bautista, Paolo Creminelli, Takeshi Kobayashi for useful discussions. 

\bibliographystyle{JHEP}
\bibliography{weyl}
\end{document}